\documentclass[11pt, prd,preprintnumbers,amsmath,amssymb, superscriptaddress]{revtex4}

\pdfoutput=1
\usepackage{latexsym}
\usepackage{amssymb}
\usepackage{epsfig,amsmath,graphics}
\usepackage{epstopdf}
\usepackage{verbatim}
\usepackage[hypertex]{hyperref}
\usepackage{graphicx}% Include figure files
\usepackage{subfigure}

\newcommand{\OO}{\mathcal{O}}

\newcommand{\aoft}{a\l t \r}
\newcommand{\boft}{b\l t \r}
\newcommand{\Trec}{T_{\text{rec}}}

\def\r{\right)}
\def\l{\left(}

\begin{document}

%\preprint{APS/123-QED}

%\title{Signatures of a Lower-Dimensional Ancestor Vacuum}
\title{Observing the Dimensionality of Our Parent Vacuum}

\author{Peter W. Graham}
%\email{pwgraham@stanford.edu}
\affiliation{Department of Physics, Stanford University, Stanford, California 94305}

\author{Roni Harnik}
%\email{roni@slac.stanford.edu}
\affiliation{Theoretical Physics Department, Fermilab, Batavia, IL60510, USA}
\affiliation{Department of Physics, Stanford University, Stanford, California 94305}

\author{Surjeet Rajendran}
%\email{surjeet@stanford.edu}
\affiliation{Center for Theoretical Physics, Laboratory for Nuclear Science and Department of Physics, Massachusetts Institute of Technology, Cambridge, MA 02139, USA}

\preprint{MIT-CTP 4119}

\date{\today}% It is always \today, today,
             %  but any date may be explicitly specified

\begin{abstract}
It seems generic to have vacua with lower dimensionality than ours.  We consider the possibility that the observable universe originated in a transition from one of these vacua.  Such a universe has anisotropic spatial curvature.  This may be directly observable through its late-time effects on the CMB if the last period of slow-roll inflation was not too long.  These affect the entire sky, leading to correlations which persist up to the highest CMB multipoles, thus allowing a conclusive detection above cosmic variance.  Further, this anisotropic curvature causes different dimensions to expand at different rates. This leads to other potentially observable signals including a quadrupolar anisotropy in the CMB which limits the size of the curvature.  Conversely, if isotropic curvature is observed it may be evidence that our parent vacuum was at least 3+1 dimensional.  Such signals could reveal our history of decompactification, providing evidence for the existence of vastly different vacua.
\end{abstract}

%\pacs{}% PACS, the Physics and Astronomy
                             % Classification Scheme.
%\keywords{Suggested keywords}%Use showkeys class option if keyword
                              %display desired
\maketitle

\tableofcontents

\section{Introduction and Summary}
\label{Sec:Intro}

Our current understanding of cosmology and high energy physics leaves many questions unanswered.  One of the most fundamental of these questions is why our universe has three large dimensions.  This may be tied to the more general question of the overall shape and structure of the universe.  In fact, it is possible that our universe was not always three dimensional or that other places outside of our observable universe have a different dimensionality.  There are surely long-lived vacua where one or more of our three dimensions are compactified, since this does not even rely on the presence of extra-dimensions and indeed happens in the Standard Model \cite{ArkaniHamed:2007gg}.  Eternal inflation can provide a means to populate these vacua, and naturally leads to a highly inhomogeneous universe on very long length scales.  Further, it seems likely that these lower-dimensional vacua are at least as numerous as three dimensional ones since there are generally more ways to compactify a greater number of spatial dimensions.  If we do indeed have a huge landscape of vacua (e.g.~\cite{Bousso:2000xa}) then it seems all the more reasonable that there should be vacua of all different dimensionalities and transitions between them (see e.g.~\cite{Krishnan:2005su, Carroll:2009dn, BlancoPillado:2009di, BlancoPillado:2009mi, Wu:2003ht}).  We will ignore the subtle issues of the likelihood of populating those vacua (the ``measure problem").  Instead we will focus on the possibility of observing such regions of lower dimensionality since surely such a discovery would have a tremendous effect on our understanding of cosmology and fundamental physics.

Our compact dimensions are generically unstable to decompactification \cite{Giddings:2004vr}.  Thus it seems possible that the universe began with all the dimensions compact (the starting point in  \cite{Brandenberger:1988aj, Greene:2009gp} for example).  In this picture our current universe is one step in the chain towards decompactifying all dimensions.  Of course, eternal inflation may lead to a very complicated history of populating different vacua, but in any case, it seems reasonable to consider the possibility that we came from a lower dimensional ``ancestor" vacuum.  We will assume that prior to our last period of slow-roll inflation our patch of the universe was born in a transition from a lower dimensional vacuum.

%However, since a longer period of slow-roll inflation requires more tuning of the inflaton potential, there may only have been just enough efolds to allow the existence of galaxies.
% since this number is physically measurable today and not dependent on the inflationary model

Our universe then underwent the normal period of slow-roll inflation.  For our signals to be observable we will assume that there were not too many more than the minimal number of efolds of inflation necessary to explain the CMB sky.  This may be reasonable because this is very near a catastrophic boundary: large scale structures such as galaxies would not form if inflation did not last long enough to dilute curvature sufficiently \cite{Vilenkin:1996ar, Garriga:1998px, Freivogel:2005vv}.  Since achieving slow-roll inflation is difficult and the longer it lasts the more tuned the potential often is, there may be a pressure to be close to this lower bound on the length of inflation.  We will actually use the energy density in curvature, $\Omega_k$, in place of the number of efolds of inflation.  The observational bound requires that $\Omega_k \lesssim 10^{-2}$ today (this corresponds to $\sim 62$ efolds for high scale inflation).  The existence of galaxies requires $\Omega_k \lesssim 1$ today (corresponding to $\sim 59.5$ efolds if we use the bound from \cite{Freivogel:2005vv}).  Thus $\Omega_k$ may be close to the observational bound today.  Other, similar arguments have also been made for a relatively large curvature today \cite{Bozek:2009gh}.

Most signals of the presence of other vacua, e.g.~bubble collisions  \cite{Aguirre:2007an, Chang:2007eq, Chang:2008gj}, also rely on this assumption.  These signals have also mostly been explored assuming that the other vacua are all 3+1 dimensional.  While an important first step, this seems like a serious oversimplification.  We find interesting differences in the case that our parent vacuum was lower dimensional.  In particular, our universe can be anisotropic, with different spatial curvatures in the different directions.  This anisotropic curvature dilutes exponentially during inflation, making the universe appear very isotropic at early times.  However, this curvature ($\Omega_k$) grows at late times, leading to several observable effects.  This anisotropic curvature sources an anisotropy in the Hubble expansion rate, since the different dimensions expand at different rates.  The most interesting signal is an anisotropy in the normal CMB curvature measurement.  The angular size of a ``standard ruler" now appears to depend on the orientation of that ruler.  In the CMB this shows up as unexpected correlations between modes of all angular sizes.  Unlike the normal curvature measurement, this anisotropic curvature measurement is not degenerate with the scale factor expansion history and is thus easier to measure.  This anistropic curvature also leads to a significant quadrupolar anisotropy in the CMB which constrains the size of $\Omega_k$.  There are possibly other observables from 21 cm measurements, direct measurements of the Hubble expansion (e.g.~from supernovae), or from searches looking for nontrivial topology of the universe.

\section{The Anisotropic Universe}
\label{Sec: metric}

In this section we will compute the evolution of a universe that began with one or two of our three spatial dimensions compactified.

\subsection{The Initial Transition}
We will consider the possibility that our universe began in a lower-dimensional vacuum.  In particular we assume that just prior to our recent period of slow-roll inflation, the currently observable part of the universe (our ``pocket universe" in landscape terminology) was in a vacuum with only one or two large, uncompactified spatial dimensions.  The other dimensions, including the one or two that will eventually become part of our three large spatial dimensions, are compactified and stable.  The universe then tunnels, nucleating a bubble of our vacuum in which three spatial dimensions are uncompactified and thus free to grow with the cosmological expansion.  We will consider starting from either a 1+1 or 2+1 dimensional vacuum.  We will not consider the 0+1 dimensional case in great detail, as it is significantly different \cite{ArkaniHamed:2007gg}.  However it is possible that it will have the same type of signatures as we discuss for the other two cases, depending on the details of the compactification manifold.

Consider first the case that the universe is initially 2+1 dimensional, and in the tunneling event one of the previously compactified spatial dimensions becomes decompactified, losing whatever forces were constraining it and becoming free to grow (in the tunneling event it may also grow directly).  We can think of this as a radion for that dimension which is initially trapped in a local minimum, tunneling to a section of its potential where it is free to roll.  Of course, the tunneling event may actually be due to a change in the fluxes wrapping the compact dimension, or in general to a change in whatever is stabilizing that dimension.  The exact nature of this tunneling will not concern us since the further evolution of the universe is relatively insensitive to this.  In all cases a bubble of the new vacuum is formed in the original 2+1 dimensional space.  The bubble wall (which is topologically an $S^1$ not an $S^2$) expands outward.  The interior of this Coleman-De Luccia bubble \cite{Coleman:1980aw} is an infinite, open universe with negative spatial curvature (see e.g.~\cite{Bousso:2005yd} for this bubble in arbitrary dimensionality space-times).  But this negative spatial curvature is only in two dimensions.  The third, previously small, dimension may be topologically an $S^1$ or an interval, but in any case will not have spatial curvature.  Thus the metric after the tunneling inside the bubble is
\begin{equation}
\label{eqn:anisotropic metric}
ds^2 = dt^2 - a(t)^2 \left( \frac{dr^2}{1-k r^2} + r^2 d \phi^2 \right) - b(t)^2 dz^2
\end{equation}
where $z$ is the coordinate of the previously compactified dimension and $k=-1$ for negative spatial curvature in the $r-\phi$ plane.  This is known as a Bianchi III spactime.

If instead the universe is initially 1+1 dimensional and two spatial dimensions decompactify in the transition then the situation will be reversed.  The single originally large dimension (now labelled with coordinate $z$) will be flat but the other two dimensions may have curvature (either positive or negative).  For example, if they were compactified into an $S^2$ they would have positive curvature and so would be described by Eqn.~\eqref{eqn:anisotropic metric} with $k=+1$, known as a Kantowski-Sachs spacetime.  Or if those two dimensions were a compact hyperbolic manifold, for example, they would be negatively curved with $k=-1$.  In fact, generically compactifications do have curvature in the extra dimensions (see for example \cite{Silverstein:2004id}).  Of course it is also possible that the two compact dimensions had zero spatial curvature.  We will not consider this special case in great detail since it does not lead to most of our observable signals.

\subsection{Evolution of the Anisotropic Universe}
We will thus assume that our universe begins with anisotropic spatial curvature, with metric as in Eqn.~\eqref{eqn:anisotropic metric}.  Immediately after the tunneling event the universe is curvature dominated, though in this case of course the curvature is only in the $r-\phi$ plane.  We assume the universe then goes through the usual period of slow-roll inflation, with a low number of efolds $\lesssim 70$ near the curvature bound.

The equations of motion (the ``FRW equations") are:
\begin{eqnarray}
\label{eqn: FRW eqn1}
\frac{\dot{a}^2}{a^2} + 2 \frac{\dot{a}}{a} \frac{\dot{b}}{b} + \frac{k}{a^2} & = & 8 \pi G \rho \\
\label{eqn: FRW eqn2}
\frac{\ddot{a}}{a} + \frac{\ddot{b}}{b} +  \frac{\dot{a}}{a} \frac{\dot{b}}{b} & = & - 8 \pi G p_r \\
\label{eqn: FRW eqn3}
2 \frac{\ddot{a}}{a} + \frac{\dot{a}^2}{a^2} + \frac{k}{a^2} & = & - 8 \pi G p_z
\end{eqnarray}
where the dot $\dot{\,}$ denotes $\frac{d}{dt}$, $\rho$ is the energy density, and $p_r$ and $p_z$ are the pressures in the $r$ and $z$ direction, i.e.~the $rr$ and $zz$ components of the stress tensor $T^\mu_\nu$.  These can be rewritten in terms of the two Hubble parameters $H_a \equiv \frac{\dot{a}}{a}$ and $H_b \equiv \frac{\dot{b}}{b}$ as
\begin{eqnarray}
\label{eqn: FRW H eqn1}
H_a^2 + 2 H_a H_b + \frac{k}{a^2} & = & 8 \pi G \rho \\
\label{eqn: FRW H eqn2}
\dot{H_a} + H_a^2 + \dot{H_b} + H_b^2 + H_a H_b & = & - 8 \pi G p_r \\
\label{eqn: FRW H eqn3}
2 \dot{H_a} + 3 H_a^2 + \frac{k}{a^2} & = & - 8 \pi G p_z
\end{eqnarray}

At least in the case of tunneling from 2+1 to 3+1 dimensions, immediately after the tunneling event the universe is curvature dominated.  In this case Eqn.~\eqref{eqn: FRW eqn3} can be solved for $a$ directly.  Since this is just the usual isotropic FRW equation, the solution is as usual $a(t) \sim t$, where $t = 0$ is the bubble wall.  Actually, since we will assume the universe transitions to a period of slow-roll inflation after curvature dominance, we will assume there is a subdominant vacuum energy during the period of curvature dominance.  This then gives a perturbutive solution accurate up to linear order in the vacuum energy $\Lambda$ of $a(t) \approx t \left( 1 + \frac{4 \pi}{9} G \Lambda t^2 \right)$.  Then we can solve Eqn.~\eqref{eqn: FRW eqn1} perturbatively for $b(t)$.  There are several possible solutions but these are reduced because we will assume that immediately after the tunneling event $\dot{b} = 0$.  If we imagine the transition as a radion field tunneling through a potential barrier then we know that the radion generically starts from rest after the tunneling.  With this boundary condition the solution to linear order in the vacuum energy is $b(t) \approx b_i \left( 1 + \frac{4 \pi}{3} G \Lambda t^2 \right)$ where $b_i$ is the initial value of $b$.  Since the period of curvature dominance ends when $t^2 G \Lambda \sim 1$, we see that roughly $a$ expands linearly while $b$ remains fixed during this period.  Thus the different expansion rates $H_a$ and $H_b$ remain very different during this period.  $H_a$ is large while $H_b \approx 0$.  The flat dimension will not begin growing rapidly until inflation begins.  At that point though, it will be driven rapidly towards the same expansion rate as the other dimensions, $H_a \approx H_b$, as we will now show.

Since our observed universe is approximately isotropic, we will only need to solve these equations in the limit of small $\Delta H \equiv H_a - H_b$.  We will always work to linear order in $\Delta H$.  Subtracting Eqn.~\eqref{eqn: FRW H eqn2} - Eqn.~\eqref{eqn: FRW H eqn3} gives
\begin{equation}
\label{eqn: delta H}
\frac{d}{dt} \Delta{H} + 3 H_a \, \Delta H + \frac{k}{a^2} = 8 \pi G \left( p_r - p_z \right) \approx 0
\end{equation}
Note that we have taken the pressure to be isotropic, $p_r = p_z \equiv p$, which is approximately true in all cases of interest to us.  This is clearly true during inflation.  During radiation dominance (RD) the radiation is in thermal equilibrium.  Since the reactions keeping it in equilibrium have rates much higher than the Hubble scales during this time, the pressure is kept locally isotropic.  During matter dominance (MD) the pressure is zero to leading order.  The sub-leading order piece due to the photons will also remain isotropic until after decoupling since the photons remain in equilibrium until this time.  After decoupling the energy density in radiation is quite small compared to the matter density.  Further, this small pressure only develops anisotropy due to the differential expansion (and hence redshifting) between the $r$ and $z$ directions.  Thus the anisotropy in pressure is proportional to both $\Delta H$ and the small overall size of the pressure and is therefore negligible for us.

The anisotropic spatial curvature in the metric \eqref{eqn:anisotropic metric} is the only effect breaking isotropy in this universe and thus the only reason for a differential expansion rate $\Delta H$.  In fact, as we will see shortly, the differential expansion $\Delta H$ is proportional to $\Omega_k$, the curvature energy density, defined to be
\begin{equation}
\Omega_k \equiv \frac{\frac{k}{a^2}}{H_a^2}.
\end{equation}
Since $\Omega_k$ grows during RD and MD and it is $\ll 1$ today, it was quite small during the entire history of the universe after the period of curvature dominance (to which we will return later).  So we will treat both $\Delta H$ and $\Omega_k$ as our small parameters and work to linear order in each.

If we combine Eqns.~\eqref{eqn: FRW H eqn3} and \eqref{eqn: delta H} we find an equation for $H_b$ which is true in the limit of small $\Delta H$
\begin{equation}
2 \dot{H_b} + 3 H_b^2 - \frac{k}{a^2} = - 8 \pi G p.
\end{equation}
Notice that this is exactly the same as the equation for $H_a$ (Eqn.~\eqref{eqn: FRW H eqn3}) but with the sign of the curvature term flipped.  Eqn.~\eqref{eqn: FRW H eqn3} for $H_a$ is just the usual isotropic FRW equation.  Thus $a(t)$ behaves exactly as it would in the normal isotropic universe with a subleading curvature component and $b(t)$ behaves as if it was the scale factor in a universe with an equal magnitude but opposite sign of curvature.

Eqn.~\eqref{eqn: delta H} can be solved easily because we only need the leading order behavior of $a$ and $H_a$ which are just the usual isotropic FRW solutions as can be seen easily since Eqns.~\eqref{eqn: FRW eqn3} and \eqref{eqn: FRW H eqn3} are just the usual FRW equations.  Solving Eqn.~\eqref{eqn: delta H} during the eras of interest and keeping only the inhomogeneous solutions yields
\begin{eqnarray}
\text{Inflation} \phantom{xRD} &  \frac{\Delta H}{H_a}  & = - \Omega_k \\
\text{RD} \phantom{xxxRD} &  \frac{\Delta H}{H_a}  & = - \frac{1}{3} \Omega_k \\
\text{MD} \phantom{iiixRD} &  \frac{\Delta H}{H_a} & = - \frac{2}{5} \Omega_k
\end{eqnarray}
As we will show later, the homogeneous solutions all die off as faster functions of time and are thus negligible.  Interestingly, this implies that $\Delta H$ is effectively independent of initial conditions.  At every transition some of the homogeneous solution for $\Delta H$ is sourced, for example to make up the missing $- \frac{2}{3} \Omega_k$ when transitioning from inflation to RD.  But this homogeneous piece dies off faster, leaving only the inhomogeneous piece which is independent of the initial value of $\Delta H$.

To find the solutions for the scale factors $a(t)$ and $b(t)$ up to linear (subleading) order in the curvature, we solve Eqns.~\eqref{eqn: FRW eqn1}, \eqref{eqn: FRW eqn2}, and \eqref{eqn: FRW eqn3} perturbatively in $\Omega_k$.  The leading order behavior comes from the dominant energy density (vacuum energy, radiation, or matter in our three eras).  We will only need the solution during MD so we can assume $p_r = p_z = 0$ then.  Eqn.~\eqref{eqn: FRW eqn3} contains no $b$'s so it can be solved directly for $a(t)$.  Once we have the solution for $a(t)$ up to linear order in $\Omega_k$ we then plug in to Eqn.~\eqref{eqn: FRW eqn2} to find $b(t)$ also to linear order.  The solutions during MD to linear order in $\Omega_k$ are
\begin{eqnarray}
a(t) = c_0 \, t^\frac{2}{3} \left( 1 - \frac{9 k}{20 c_0^2} \, t^\frac{2}{3} \right) \approx  c_0 \, t^\frac{2}{3} \left( 1 - \frac{\Omega_k}{5} \right) \\
b(t) = c_0 \, t^\frac{2}{3} \left( 1 + \frac{9 k}{20 c_0^2} \, t^\frac{2}{3} \right) \approx  c_0 \, t^\frac{2}{3} \left( 1 + \frac{\Omega_k}{5} \right)
\end{eqnarray}
where $c_0$ is an arbitrary, physically meaningless constant arising from the coordinate rescaling symmetry.

Thus this universe always has a differential expansion rate between the $z$ direction and the $r$-$\phi$ directions which is proportional to $\Omega_k$.  The precise constant of proportionality depends only on the era (inflation, RD, or MD) and not on initial conditions.  Further, the $r$-$\phi$ plane expands as in the usual isotropic FRW universe, while the $z$ direction expands as if it was in that same universe except oppositely curved.
During an initial period of curvature dominance the $z$ dimension remains constant while the other two dimensions expand, diluting curvature.  During this period the expansion rates $H_a$ and $H_b$ are maximally different.  Then a period of slow-roll inflation takes over.  During this period the expansion rates are driven exponentially close together.  This difference in expansion rates is largest at the beginning of inflation, immediately after curvature dominance, when $\Omega_k$ is still large.  During inflation curvature dilutes exponentially as $\Omega_k \propto a^{-2}$.  So at the end of inflation the differential expansion rate is completely negligible $\frac{\Delta H}{H} \sim e^{-60}$.  Then during RD $\Omega_k$ and hence also $\Delta H$ remain small, though growing as $\propto a^2$.  During MD $\Omega_k$ and $\Delta H$ continue to grow $\propto a$, finally reaching their maximal value when the universe transitioned to vacuum energy dominance around redshift $\sim 2$.  Since this final transition was so recent (and the homogeneous solution for $\Delta H$ has not even had much time to die off yet) we will approximate the universe as matter dominated until today.

\section{Observables}
\label{Sec: Observables}
%The universe created in the scenarios described in Section \ref{Sec: metric} are born with anisotropic curvature. This anisotropy is initially exponentially diluted by inflation, but then its fractional contribution to the energy density of the universe grows back during the epochs of radiation and matter domination.
In this section, we discuss the late time observables of anisotropic curvature. We begin by computing its effects on standard rulers.  These effects emerge due to the warping of null geodesics in the anisotropic background metric. Null geodesics along different directions are warped differently by the curvature, leading to differences in the observed angular size of standard rulers in the sky. Following this discussion, we compute the effect of anisotropic curvature on the CMB. The CMB is also affected by the warping of the null geodesics that propagate from the surface of last scattering to the current epoch. This warping affects the relation between the angle at which a CMB photon is observed today and the point at which it was emitted during recombination. In addition to this effect, the anisotropic metric discussed in Section \ref{Sec: metric} also leads to differential Hubble expansion. This leads to an anisotropic red shift in the universe, which causes a late time observer to see additional temperature anisotropies in the CMB. We conclude the section with a discussion of additional measurements that could be performed in upcoming experiments. 

\subsection{Standard Rulers}
\label{SubSec:Rulers}

In this Section we present a calculation of the effect on standard rulers.  While this is not a directly observable effect itself since we have no exact standard rulers in the sky, it does provide good intuition for the following calculation of the actual CMB observables in Section \ref{SubSec:CMB}.  Further, many of the results of this section are used directly in that calculation.

The spacetimes \eqref{eqn:anisotropic metric} considered in this paper are curved and anisotropic. A canonical method to observe curvature is through the measurement of the angular sizes of standard rulers. Curvature modifies the Euclidean relationship between the measured angle and the linear size of the ruler. In a universe with anisotropic curvature, we expect this deviation from Euclidean geometry to change with the angular position and orientation of the ruler. 

Motivated by the use of baryon acoustic oscillations as cosmological standard rulers, we compute the present day angular size of standard rulers located at the surface of last scattering. This calculation gives intuition for the effects of anisotropic curvature on the CMB (studied in detail in section \ref{SubSec:CMB}). To do so, we first determine the null geodesics that connect the surface of last scattering to a present day observer. The angle subtended between the two null geodesics that reach the end points of the standard ruler is then the angular size of the ruler. For simplicity, we assume that the universe was matter dominated throughout the period between recombination and the present epoch.  

We work with the metric 
\begin{eqnarray}
ds^{2} & = & dt^{2} - \aoft^{2} \l dr^{2} + \sinh^{2}\l r \r d\phi^2 \r - \boft^2 dz^2
\label{Eqn:sinhmetric}
\end{eqnarray}
which is produced  when a $3+1$ dimensional universe is produced by tunneling from a $2+1$ dimensional vacuum. We restrict our attention to this scenario in order to facilitate concrete computation. However, our results can be applied to a wide class of  scenarios that lead to anisotropic geometries. The metric  \eqref{Eqn:sinhmetric} describes a universe where  two of the spatial dimensions (parameterized by the coordinates $\l r, \phi \r$ in \eqref{Eqn:sinhmetric}) have negative curvature and grow with scale factor $\aoft$. The other dimension, parameterized by the coordinate $z$ in \eqref{Eqn:sinhmetric}, grows with scale factor $\boft$. The space-time geometry of such a universe can also be described using the metric  \eqref{eqn:anisotropic metric} with $k = -1$. These metrics are related by a coordinate transformation and they yield identical FRW equations \eqref{eqn: FRW eqn1}, \eqref{eqn: FRW eqn2} and \eqref{eqn: FRW eqn3} for the scale factors $\aoft$ and $\boft$. Th
 is setup also describes anisotropic universes with positive curvature (equation \eqref{eqn:anisotropic metric} with $k=+1$). Such a universe is described by the metric \eqref{Eqn:sinhmetric} with the $\sinh^{2}\l r \r$ term replaced by $\sin^{2}\l r \r$. With this metric, the FRW (equations \eqref{eqn: FRW eqn1}, \eqref{eqn: FRW eqn2} and \eqref{eqn: FRW eqn3}) and null geodesic equations \eqref{Eqn:Geodesics} have the same parametric forms. Our calculations also apply to this case, with the difference between the two cases being captured by the sign of the curvature term $\Omega_k$. 

\begin{figure}
\begin{center}
\includegraphics[width = 4.0 in]{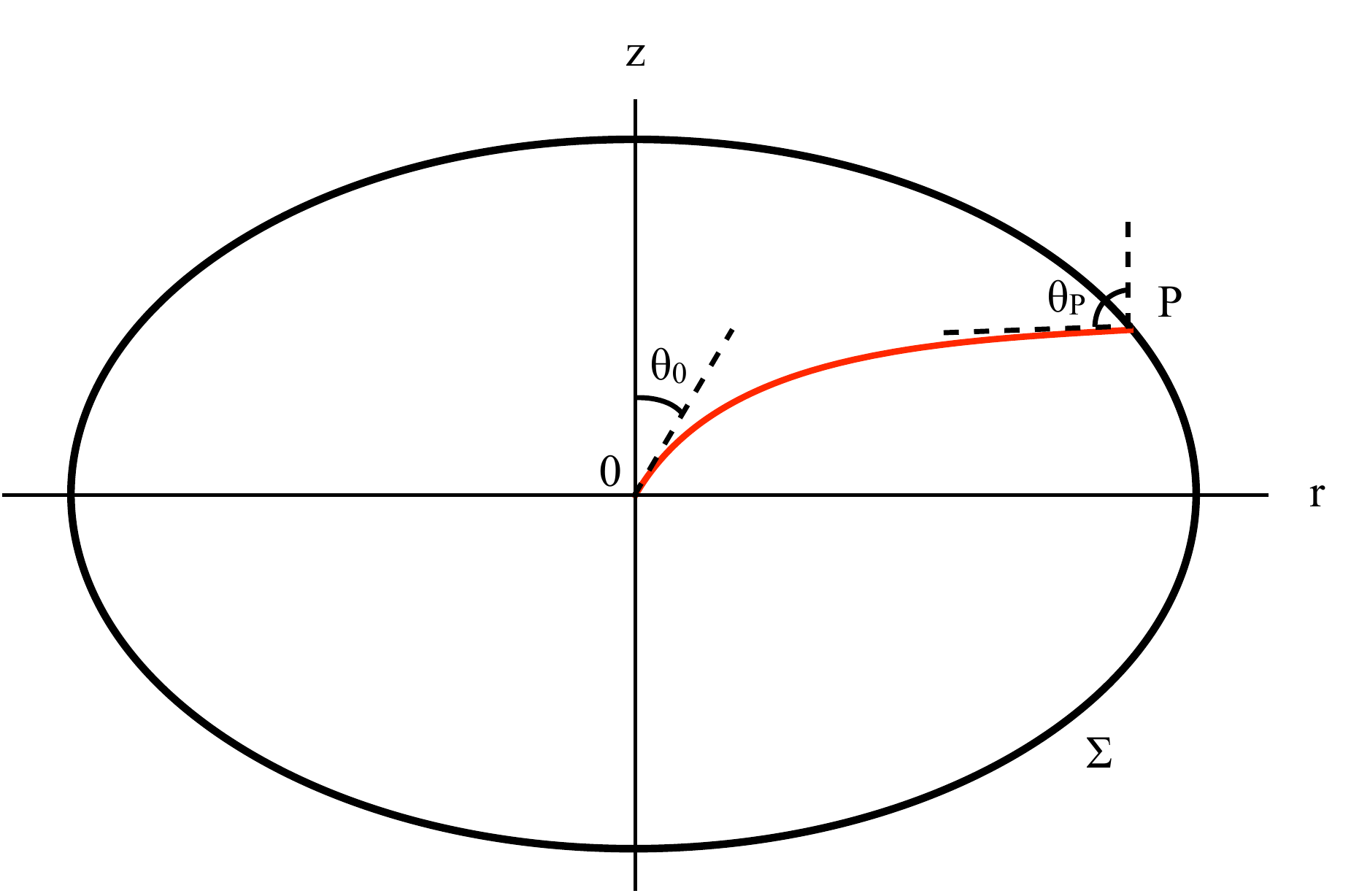}
\caption{ \label{Fig:geodesics} A depiction of the motion of a photon (red curve) from a point P on  the surface of last scattering $\Sigma$ (black ellipse) to an observer O. Without loss in generality, the observer's position can be taken as the origin of the coordinate system. The anisotropic curvature causes $\Sigma$ to deviate from sphericity and warps the photon trajectories.  $\theta_0$ is the angle between the photon's trajectory and the observer O's $z$ axis. $\theta_P$ is the angle  between the photon's trajectory and the $z$ axis at $P$.}
\end{center}
\end{figure}
 
 An observer O (see figure \ref{Fig:geodesics}) at the present time receives  photons from the surface of last scattering $\Sigma$. This photon follows a null geodesic. In computing this null geodesic, we can assume without loss in generality that the point O lies at the origin of the coordinate system. With this choice, we focus on geodesics that lie along a direction of constant $\phi$. These geodesics contain all the information required to describe our setup. The geodesics that connect the point O with the surface of last scattering have zero velocity along the $\phi$ direction. The $O\l 2 \r$ symmetry in the $\l r, \phi \r$ plane then implies that $\phi$ remains constant during the subsequent evolution of the geodesic. Using the metric \eqref{Eqn:sinhmetric}, the null geodesic equations that describe the photon's trajectory $\l r\l t \r, z \l t \r \r$ are 
\begin{eqnarray}
\nonumber
\ddot{r} + \dot{r} H_a \l 1 + \frac{\Delta H}{H_a} \l 1 - \dot{r}^2 \aoft^2 \r \r& =&0  \\
\ddot{z} + \dot{z} H_b \l 1 - \frac{\Delta H}{H_b} \l 1 - \dot{z}^2 \boft^2 \r \r& = &0 
\label{Eqn:Geodesics}
\end{eqnarray}
where the dots denote derivatives with respect to $t$. With the boundary condition that the null geodesic reaches O at time $t_0$, equation \eqref{Eqn:Geodesics} can be solved perturbatively to leading order in $\Omega_k$. The coordinates $\l r_P, z_P \r$ on $\Sigma$ from which the photon is emitted are 
\begin{eqnarray}
\nonumber
r_P &=& \sin \alpha \,  \frac{3 \,  t_0^{\frac{1}{3}}}{c_0} \,  \l 1 - \frac{\Omega_{k_0}}{3}\l \frac{4}{5}  +  \cos 2 \alpha \r \r \\
z_P &=& \cos \alpha \,  \frac{3 \,  t_0^{\frac{1}{3}}}{c_0} \,  \l 1 - \frac{\Omega_{k_0}}{3}\l -\frac{4}{5}  + \cos 2 \alpha \r \r
\label{Eqn:Coordinates}
\end{eqnarray}
In the above expression, $\alpha$ is a parameter that governs the direction in the $\l r, z \r$ plane from which the photon is received at O and  $\Omega_{k_0}$ denotes the fractional energy density in curvature at the present time. The physical angle $\theta$ between the photon's trajectory and the $z$ axis, as measured by a local observer, is different from $\alpha$, and is given by 
\begin{eqnarray}
\tan \l\theta\r &=& \l \frac{\aoft}{\boft} \frac{dr}{dz}\r + \OO \l H_a\r
\label{Eqn:thetadefinition}
\end{eqnarray}
For the geodesics computed in \eqref{Eqn:Coordinates}, the relation between the parameter $\alpha$ and the physical angle $\theta_0$ observed at O is  
\begin{eqnarray}
\tan \l\theta_0\r & = &\l 1 - \frac{2}{5} \Omega_{k_0}\r \, \tan \alpha
\end{eqnarray}
The $\OO \l H_a \r$ corrections in the definition \eqref{Eqn:thetadefinition} arise because the coordinates $\l t, r, \phi, z \r$ used to describe the metric \eqref{Eqn:sinhmetric} are not locally flat. Local coordinates $\l \tilde{t}, \tilde{r}, \tilde{\phi}, \tilde{z} \r$ can be constructed at any point $\l t_Q, r_Q, \phi_Q, z_Q \r$ of the space-time. These two sets of coordinates are related by 
\begin{eqnarray}
\nonumber
t &  = & t_Q + \tilde{t} - \frac{1}{2}\l \aoft^{2} H_a  \l \tilde{r}^2 + \sinh^{2}\l r_Q \r \, \tilde{\phi}^{2} \r  + \boft^2 H_b \tilde{z}^2  \r \\ 
\nonumber
r & = & r_Q + \tilde{r} - \frac{1}{2} \l 2 \, H_a \,  \tilde{r} \,  \tilde{t} \, - \,  \cosh \l r_Q\r \, \sinh \l r_Q\r \, \tilde{\phi}^2\r \\ 
\nonumber
\phi & = & \phi_Q  + \tilde{\phi} - \tilde{\phi} \l   H_a  \, \tilde{t}  \, + \, \coth \l r_Q\r \, \tilde{r}   \r \\
z & = & z_Q + \tilde{z} \l 1 - H_b \, \tilde{t} \r
\label{Eqn:LocalCoordinates}
\end{eqnarray}
The coordinate transformations in \eqref{Eqn:LocalCoordinates} imply that operators constructed from global coordinates ({\it e.g.} $\frac{d}{dr}$) differ from the corresponding operator in the local inertial frame ({\it e.g.} $\frac{d}{d\tilde{r}}$) by quantities $\sim \OO \l H_a \tilde{r}\r$.  The difference between these operators is suppressed by the ratio of the size of the local experiment over the Hubble radius. These differences are negligible for any local experiment today. The angle defined by \eqref{Eqn:thetadefinition} is therefore very close to the physical angle measured by a local experiment and we will use this definition for subsequent calculations.

\begin{figure}
\begin{center}
\includegraphics[width = 4.5 in]{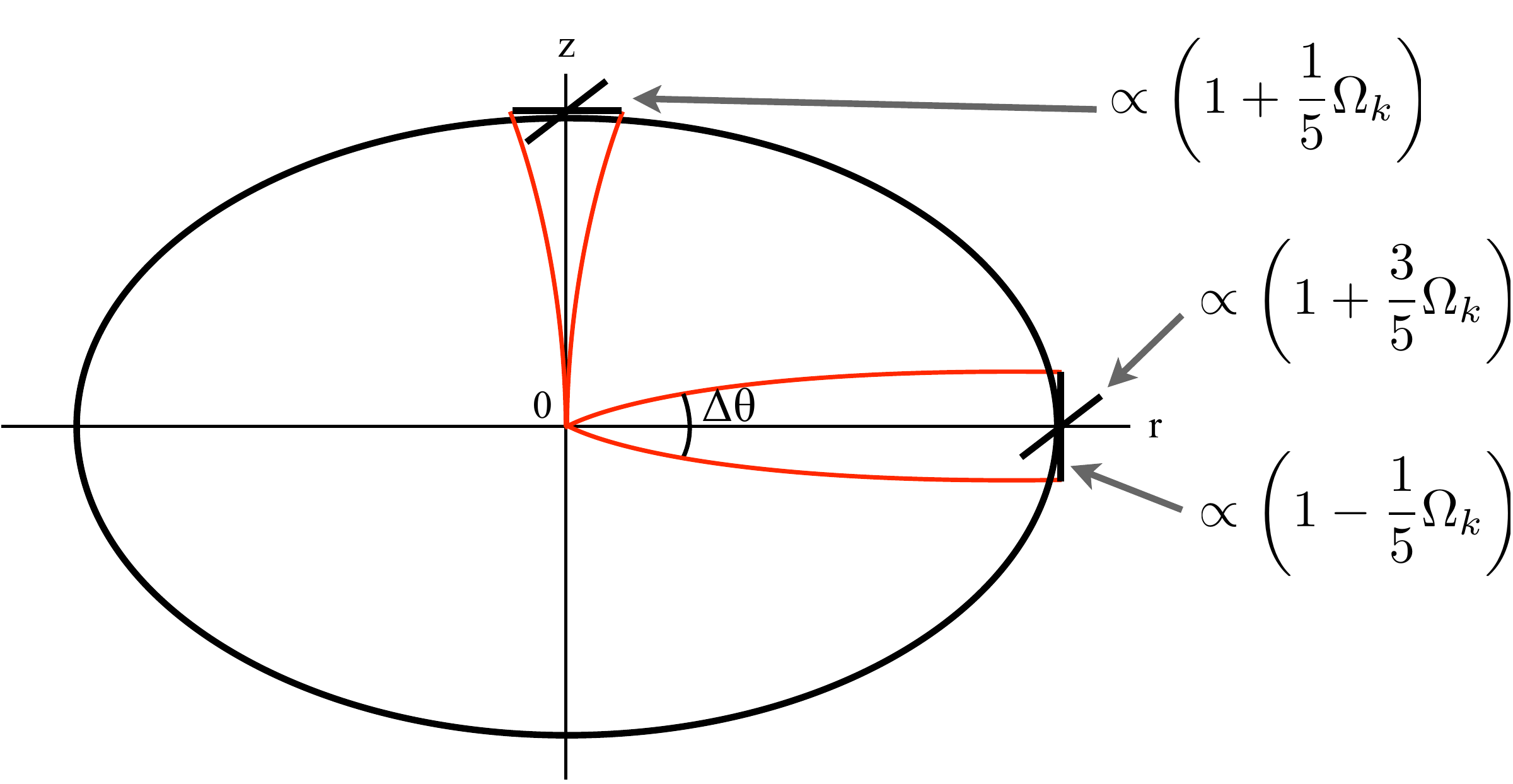}
\caption{ \label{Fig:rulers} The effect of the anisotropic curvature on a measurement of the angular size of standard rulers. The black ellipse is the surface of recombination and the red lines are photon paths from standard rulers on this surface to the observer at O. The standard rulers are depicted by the thick straight lines. The angular size varies depending upon the location and orientation of the ruler.}
\end{center}
\end{figure}

With the knowledge of the geodesics \eqref{Eqn:Coordinates}, we can calculate the angular size of a standard ruler of length $\Delta L$ at the time of recombination. Since $\Omega_k$ is very small during this time, the physical size of the ruler is independent of its location and orientation.  First,  consider a ruler oriented in the $z$ direction. This ruler lies between the co-ordinates $\l r \l t_{r} \r, z \l t_r \r \r $ and $\l r \l t_{r} \r  + \Delta r, z \l t_r \r + \Delta z\r$ at the time $t_r$ of recombination. The length of this ruler is 
\begin{eqnarray}
\l a\l t_r \r \Delta r \r^{2} + \l b \l t_r \r \Delta z \r^{2} & = & \Delta L ^2
\label{Eqn:RulerSize}
\end{eqnarray}
Using  \eqref{Eqn:Coordinates}  and \eqref{Eqn:thetadefinition} in \eqref{Eqn:RulerSize}, we find that the angular size $\Delta \theta$ subtended by a ruler of length $\Delta L$ at a local experiment O is 
\begin{eqnarray}
\Delta \theta \l \theta \r & = & \frac{\Delta L}{3 \, t_r^{\frac{2}{3}} \, t_0^{\frac{1}{3}}} \l 1 + \frac{\Omega_{k_0}}{5} \cos 2 \theta \r
\label{Eqn:AngleChange}
\end{eqnarray}
A similar procedure can also be adopted to describe standard rulers that lie along the $\l r, \phi \r$ plane. The angular size of these rulers is given by the angle $\phi$ between the  null geodesics that connect the ends of the ruler to the origin. Following the above procedure, this angular size $\Delta \phi$ is 
\begin{eqnarray}
\Delta \phi & = & \frac{\Delta L}{3 \, t_r^{\frac{2}{3}} \, t_0^{\frac{1}{3}}} \l 1 + \frac{3}{5} \, \Omega_{k_0}\r
\label{Eqn:rphiangle}
\end{eqnarray}
The angular size of a standard ruler thus changes when its location and orientation are changed (see figure \ref{Fig:rulers}). For a ruler located at $z = 0$ ({\it i.e.} $\theta  = \frac{\pi}{2}$) the warp of the angle (in equations \eqref{Eqn:AngleChange} and \eqref{Eqn:rphiangle}) changes from $\l 1 + \frac{3}{5} \, \Omega_{k_0}\r$ for a ruler in the $\l r, \phi \r$ plane to $\l 1 - \frac{1}{5} \, \Omega_{k_0}\r$ for a ruler oriented in the $z$ direction. Similarly, as a ruler  oriented in the $z$ direction is moved from $\theta_0 = 0$ to $\theta_0 = \frac{\pi}{2}$, the angular warp factor changes from $\l 1 - \frac{1}{5} \, \Omega_{k_0}\r$  to $\l 1 +  \frac{1}{5} \, \Omega_{k_0}\r$.  The reason for this change can be traced to the fact that for a ruler oriented in the $z$ direction, all of the angular warp occurs due to the effect of the curvature on the scale factor.  $\aoft$ and $\boft$ expand as though they have the same magnitude of the curvature but with opposite sig
 n. Consequently,  the angular warps along the two directions also have the same magnitude, but are of opposite sign. This angular dependence is an inevitable consequence of the anisotropic curvature $\Omega_k$ endemic to this metric. We  note that this measurement of the anisotropic curvature is  relatively immune to degeneracies from the cosmological expansion history since the angular size changes depending upon the orientation of the ruler along every line of sight. We discuss how this measurement can be realized using CMB measurements in section \ref{SubSec:StatisticalAnisotropy}.

\subsection{Effect on the CMB}
\label{SubSec:CMB}
The CMB offers a unique probe of the space-time geometry between the surface of last scattering  and the current epoch. The spectral characteristics of the CMB photons at the time of last scattering are well determined. Differences between this well determined spectrum and observations of the local flux of CMB photons arise during the propagation of the photons from recombination to the present epoch.  These differences can be used to trace the space-time geometry since these photons travel along null geodesics  of the geometry. In this section, we use the trajectories of CMB photons computed in sub section \ref{SubSec:Rulers} to derive the spectrum of the CMB flux observed today. 

The CMB flux observed at O (see figure \ref{Fig:geodesics}) is 
\begin{eqnarray}
\Phi_0 \l E_0 \r & = & \frac{dN_0\l E_0 \r}{\sin\theta_0 d\theta_0 d\phi_0 dA_0 dt_0 dE_0}
\label{Eqn:flux}
\end{eqnarray}
where $dN_0\l E_0 \r$ is the number of photons with energies between $E_0$ and $E_0 + dE_0$ received at O within a solid angle $\sin\theta_0 d\theta_0 d\phi_0$ in an area $dA_0$ during a time $dt_0$. The angle $\theta_0$ is defined as per \eqref{Eqn:thetadefinition} since that definition corresponds to the physical angle that a local observer measures between the photon's trajectory and the $z$ axis. The photons that are received at O within this solid angle were emitted from the point P on the surface of last scattering $\Sigma$ (see figure \ref{Fig:geodesics}). Since the geometry of the universe (equation \eqref{Eqn:sinhmetric}) is curved, the solid angle $\sin\theta_P d\theta_P d\phi_P$  is different from the solid angle at O. The energy $E_P$ at which the photon is emitted  is also different from the energy $E_0$ at which it is received owing to the expansion of the universe. Furthermore, due to the differential expansion of the $\l r, \theta \r$ plane and the $z$ directi
 on, this energy shift is also a function of the solid angle. The photons received in the space-time volume $dA_0\, dt_0$ are emitted from a volume $dA_P \, dt_P$. The ratio of these volume elements is proportional to the expansion of the universe. Incorporating these effects, the flux \eqref{Eqn:flux} can be expressed as 
\begin{eqnarray}
\Phi_0 \l E_0 \r & = & \frac{dN_P\l E_P \r}{\sin\theta_P d\theta_P d\phi_P dA_P dt_P dE_P}  \l\frac{\sin\theta_P d\theta_P d\phi_P}{\sin\theta_0 d\theta_0 d\phi_0}\r \l\frac{dA_P dt_P}{dA_0 dt_0}\r \l\frac{dE_P}{dE_0}\r
\label{Eqn:fluxcalculus}
\end{eqnarray}
or, in terms of the emission flux $\Phi_P$, 
\begin{eqnarray}
\Phi_0 \l E_0 \r & = & \Phi_P \l E_P \r  \l\frac{\sin\theta_P d\theta_P d\phi_P}{\sin\theta_0 d\theta_0 d\phi_0}\r  \l\frac{dA_P \,dt_P}{dA_0 \, dt_0}\r \l\frac{dE_P}{dE_0}\r
\label{Eqn:fluxrelation}
\end{eqnarray}

To find the local flux, we have to relate the geometric and energy elements in \eqref{Eqn:fluxrelation} at P to those at O. We begin with the angle $\theta_0$. Using the definition \eqref{Eqn:thetadefinition} of $\theta$ and the solution \eqref{Eqn:Geodesics} for the geodesic,  we solve for $\theta_0$ along the null geodesic and find that 
\begin{eqnarray}
\theta_0 &=& \theta_P + \frac{1}{5}\Omega_{k_0} \sin \l 2 \, \theta_P \r + \OO \l \Omega_{k_P}\r   + \OO \l \Omega_{k}^2\r 
\label{Eqn:thetasolution}
\end{eqnarray}
where $\theta_0$ and $\theta_P$ are the angles of the photon's trajectory at the observer's present location O and the point P  (see figure \ref{Fig:geodesics}) on the surface of last scattering which is connected to O by the null geodesic. We have ignored contributions of order $\Omega_{k_P}$, the fractional energy in curvature at the time of recombination, in this solution. This is justified since $\Omega_{k_P} \ll \Omega_{k_0}$. The angle $\phi$ is unaffected by the anisotropic curvature since there is an $O\l 2 \r$ symmetry in the $\l r, \phi \r$ plane. Consequently, $d \phi_0 = d \phi_P$. 

The volume elements are proportional to the expansion of the universe and are given by 
\begin{eqnarray}
 \frac{dA_P \,dt_P}{dA_0 \, dt_0} &=& \l \frac{a_P}{a_{0}}\r^2 \l \frac{ b_P}{ b_{0}}\r
\label{Eqn:VolumeRelation}
\end{eqnarray}
where  $\l a_P, b_P \r$ and  $\l a_0, b_0 \r$ are the scale factors at the points P and O respectively. Finally, we need to compute the relationship between the observed energy $E_0$ of the photon and the emission energy $E_P$. 

The energy E observed by a local observer at some point along the photon's trajectory is given by 
\begin{eqnarray}
E^2 &=& \l a \frac{dr}{d\tau} \r ^2 + \l b \frac{dz}{d\tau}\r^2
\end{eqnarray}
where $\tau$ is an affine parameter along the photon trajectory. Using the geodesic equations \eqref{Eqn:Geodesics} and the above expression, the present day energy $E_0$ is
\begin{eqnarray}
E_0 & = & E_P \l \frac{a_P}{a_0} \r \l 1 - \frac{2}{5} \Omega_{k_0} \cos^{2} \theta_P\r 
\label{Eqn:EnergyRelation}
\end{eqnarray}
Incidentally, this expression can also be arrived at by red shifting the momentum components of the photon along the radial and $z$ directions by $\l  \frac{a_P}{a_0},  \frac{b_P}{b_0} \r$ respectively. 

We now have all the ingredients necessary to compute the present day flux $\Phi_0$ given an initial flux $\Phi_P$ at recombination. Since $\Omega_k$ prior to recombination is much smaller than $\Omega_{k_0} \ll 1$, the CMB spectrum at recombination is identical to that of the usual FRW universe. In particular, the CMB at P is a black-body at a temperature $T_P$, with its spectrum, independent of angle, given by the Planck distribution
\begin{eqnarray}
\Phi_P \l E_P \r & = & \frac{E_P^2}{\exp\l \frac{E_P}{T_P} \r - 1}
\label{Eqn:PlanckDistribution}
\end{eqnarray}

We define $\widetilde{T}_0 = T_P \l \frac{a_P^2 b_P}{a_0^2 b_0}\r^{\frac{1}{3}}$. This definition  is motivated by the fact that CMB temperature should redshift roughly as the ratio of the scale factors of expansion. In this anisotropic universe, where two dimensions expand with scale factor $a$ and the other with scale factor $b$, the quantity $\l\frac{a_P^2 b_P}{a_0^2 b_0}\r^{\frac{1}{3}}$ is roughly the mean expansion factor. Using  \eqref{Eqn:thetasolution}, \eqref{Eqn:VolumeRelation},  \eqref{Eqn:EnergyRelation}  and \eqref{Eqn:PlanckDistribution} in \eqref{Eqn:fluxrelation}, we get 
\begin{eqnarray}
\Phi_0 \l E_0, \theta_0 \r & = & \frac{E_0^2}{\exp\l \frac{E_0}{\widetilde{T}_0} \l 1 + \frac{8}{15}  \sqrt{\frac{\pi}{5}}  \Omega_{k_0} Y_{20} \l \theta_0, \phi_0 \r \r  \r - 1}
\label{Eqn:FinalDistribution}
\end{eqnarray}
where $Y_{20}\l \theta_0, \phi_0 \r$ is the spherical harmonic with $l = 2, m=0$. 

It is well known that primordial density fluctuations lead to temperature anisotropies $\sim 10^{-5}$ in the CMB. The temperature $\widetilde{T}_0$ in \eqref{Eqn:FinalDistribution} inherits these anisotropies and is consequently a function of the angle $\l \theta, \phi \r$ in the sky. Using this input, the distribution in \eqref{Eqn:FinalDistribution}  describes a blackbody with a temperature  
\begin{eqnarray}
T_0 \l \theta_0, \phi_0 \r & = &\widetilde{T}_0 \l \theta_P, \phi_P \r  \l1 - \frac{8}{15} \sqrt{\frac{\pi}{5}} \Omega_{k_0} Y_{20} \l \theta_0, \phi_0\r\r
\label{Eqn:CMBTemperature}
\end{eqnarray}
at a given direction $\l \theta_0, \phi_0 \r$ in the sky. Note that the relation between the present day temperature $T_0$ and the temperature at recombination $T_P$ is warped both by the multiplicative factor (the term in brackets) in \eqref{Eqn:CMBTemperature} as well as the difference between the angles $\l \theta_P, \phi_P \r$  and $\l \theta_0, \phi_0\r$. Both these effects are proportional to $\Omega_{k_0}$ and lead to effects in the CMB. In the following subsections, we highlight the key observables of this spectrum. 

\subsubsection{The Quadrupole}
\label{SubSec:CMBQuadrupole}
The temperature $T_0 \l \theta_0, \phi_0 \r$ is nearly uniform across the sky with an average temperature $\bar{T}_0$ and primordial temperature fluctuations $\sim 10^{-5}$. Substituting for $T_0 \l \theta_0, \phi_0 \r$ in \eqref{Eqn:CMBTemperature}, we find that the anisotropic curvature leads to a quadrupole $a_{20} \sim - \bar{T}_0 \frac{8}{15} \sqrt{\frac{\pi}{5}} \Omega_{k_0} Y^{0}_2 \l \theta_0, \phi_0\r$ (see equation \eqref{Eqn:almdefinition}) in the CMB temperature. The source of this quadrupole is the differential expansion rate of the Universe between the $\l r, \theta \r$ plane and the $z$ direction (see equation \eqref{Eqn:sinhmetric}), leading to differential red shifts along these directions. These differential red shifts lead to a quadrupolar warp of the  average temperature of the surface of last scattering. Unlike the primordial perturbations which are generated during inflation, this contribution to the quadrupole in the CMB arises from the late time emerge
 nce of the anisotropic curvature. Fractionally, the additional power due to this effect is $\sim \Omega_{k_0}$. 

Current observations from the WMAP mission constrains the quadrupolar temperature variation $\sim 10^{-5}$ \cite{Larson:2010gs}. Naively, this constrains $\Omega_{k_0} \lessapprox 10^{-5}$. However, the quadrupole that is observed in the sky is a sum of the quadrupole from the primordial density fluctuations and this additional contribution from the anisotropic curvature. It is then possible for these two contributions to cancel against each other leading to a smaller observed quadrupole.  This cancellation requires a tuning between the primordial quadrupolar density perturbation and the anisotropic curvature contribution. Additionally, this tuning can be successful only if the primordial quadrupolar perturbation is $\OO \l \Omega_{k_0}\r$. 

The primordial density fluctuations are $\sim 10^{-5}$ and it is difficult for the quadrupolar fluctuations to be much higher than this level. However, in a universe with a small number of e-foldings of inflation, the quadrupole is the mode that leaves the horizon at the very beginning of inflation and is therefore sensitive to physics in the primordial pre inflationary space-time. These phenomena are not constrained by inflationary physics and they could lead to additional power in the quadrupolar modes  \cite{Fialkov:2009xm, Chang:2008gj, Chang:2007eq}.  It is therefore possible for the power in the primordial quadrupolar mode to be somewhat larger, leading to possible cancellation of the  quadrupole from the late time anisotropic curvature.  In fact, the measured quadrupole in our universe has significantly less power than expected from a conventional $\Lambda$CDM model \cite{Larson:2010gs}. This anomaly may already be an indication of non-inflationary physics affecting th
 e quadrupole \cite{Freivogel:2005vv}. There is also some uncertainty on the overall size of the quadrupole. For example, astrophysical uncertainties \cite{Bennett:2010jb, Francis:2009pt} could potentially make the quadrupole in the CMB larger by a factor $\sim 2 - 3$. Owing to these uncertainties, it may be possible for $\Omega_{k_0}$ to be as large as  $10^{-4}$ without running afoul of observational bounds. Values of  $\Omega_{k_0}$ significantly larger than $\sim 10^{-4}$ may also be possible. However, the additional tuning required to cancel the associated quadrupole may disfavor this possibility.

It is interesting to note that  anisotropic curvature  is  much more constrained than isotropic curvature. Current cosmological measurements constrain the isotropic curvature contribution $\lessapprox 10^{-2}$ \cite{Larson:2010gs}. However, anisotropic curvature leads to temperature anisotropies in the sky.  Since these anisotropies are well constrained by current measurements, the bounds on  $\Omega_{k_0} \lessapprox 10^{-4}$ are more stringent (for example, see \cite{Demianski:2007fz}). This bound is close to the cosmic variance limit on $\Omega_{k_0} \approxeq 10^{-5}$. Consequently, there is an observational window of $10^{-5} \lessapprox \Omega_{k_0} \lessapprox 10^{-4}$ where the anisotropic curvature can be discovered.

\subsubsection{Statistical Anisotropy}
\label{SubSec:StatisticalAnisotropy}

In this subsection we discuss the effects of anisotropic curvature on the power spectrum of the CMB. The warping of  standard rulers by the anisotropic curvature (see section \ref{SubSec:Rulers}) manifests itself in the CMB through these effects. At the present time, an observer O (see figure \ref{Fig:geodesics}) characterizes the CMB through the spectrum defined by
\begin{equation}
a_{lm}=\int d\Omega\, T_0(\theta_0,\phi_0) Y_{lm}(\theta_0,\phi_0)
\label{Eqn:almdefinition}
\end{equation}
where the present day temperature $T_0$ is defined in  equation \eqref{Eqn:CMBTemperature}.  The correlation functions $\langle a_{lm} a^*_{l'm'} \rangle$ of this spectrum contain all the information  in the CMB. In a statistically isotropic universe, all non-diagonal correlators of the $a_{lm}$ vanish. Anisotropies mix different angular scales and will populate these non-diagonal correlators. We compute them in this section.

$T_0$ inherits the density fluctuations at the time of recombination.  Since anisotropies were small prior to recombination, we will assume that the spectrum of density fluctuations at recombination is given by a statistically isotropic, Gaussian distribution. The small anisotropies prior to recombination do alter this distribution and can give rise to additional observables   \cite{BlancoPillado:2010uw, Adamek:2010sg}. However, these corrections are proportional to the anisotropic curvature $\Omega_{k_r}$  during recombination   \cite{BlancoPillado:2010uw, Adamek:2010sg}.  Since $\Omega_{k_r}$ is smaller than the present day anisotropic curvature $\Omega_{k_0}$ by a factor of $\sim 1000$, the experimental observables are dominated by the effects of the late time anisotropic curvature $\Omega_{k_0}$. In order to compute these late time effects, it is sufficient to assume that the spectrum of density fluctuations at recombination is statistically isotropic and Gaussian. We will therefore make this assumption for the rest of the paper. Our task is to start with this spectrum at recombination and compute the characteristics of the CMB spectrum observed by O.

The anisotropic curvature warps the CMB spectrum at O in three ways.  First, the photons from the surface of last scattering that reach O do not lie on a spherical surface (see figure \ref{Fig:geodesics}). This warped surface $\Sigma$ is described by equation \eqref{Eqn:Coordinates}, where the deviations from sphericity are proportional to the late time curvature $\Omega_{k_0}$. Second, the angle $\theta_0$ at which the photon is received at O is different from the co-ordinate angle $\beta$ on the surface of recombination at which this photon was originally emitted. Third, the photon is red-shifted when it reaches O. This red-shift also depends upon the angle since the anisotropic curvature causes a differential Hubble expansion leading to anisotropic red-shifts. 

We first determine the spectrum on $\Sigma$, the surface from which photons at recombination reach O. $\Sigma$ can be described using spherical coordinates  $\l R, \beta, \phi \r$ . $R$ is the physical distance at recombination between $O$ and a point $P$ on $\Sigma$ (see figure \ref{Fig:geodesics}), $\beta$ is the polar angle between the $z$ axis and the unit vector at O that lies in the direction of $P$ and $\phi$ is the azimuthal angle. These flat space coordinates appropriately describe the recombination surface since the spatial curvature was very small during this period. In particular, the polar angle $\beta$ is given by 
\begin{eqnarray}
\tan \beta & = & \frac{r_P}{z_P} 
\label{Eqn:betadefinition}
\end{eqnarray}
while the physical distance $R$ (using equation \eqref{Eqn:Coordinates}) is 
\begin{eqnarray}
R\l \beta \r & = & 3 \, t_{0}^{\frac{1}{3}} \, t_r^{\frac{2}{3}} \, \l 1 + \frac{\Omega_{k_0}}{45} - \frac{8 \, \Omega_{k_0}}{45} \sqrt{\frac{\pi}{5}} Y_{20} \l \beta, \phi \r \r
\label{Eqn:Rdefinition}
\end{eqnarray}
The spectrum at $\Sigma$ can be characterized by
\begin{equation}
b_{lm}= \left(\frac{a_P^2 \, b_P}{a_0^2 \, b_0}\right)^{1/3}
 \int_{\Sigma} d\Omega\, \Trec(\beta, \phi) Y_{lm}(\beta,\phi)
\label{Eqn:blmdefinition}
\end{equation} 
where $\Trec$ is the temperature at the recombination surface. The multiplicative factor $\left(\frac{a_P^2 \, b_P}{a_0^2 \, b_0}\right)^{1/3}$ in \eqref{Eqn:blmdefinition} is introduced for convenience. It accounts for the red shift of the mean temperature from the era of recombination to the present time, but does not introduce additional correlations in the power spectrum.  With this definition of $b_{lm}$, the correlation functions of the distributions \eqref{Eqn:almdefinition} and \eqref{Eqn:blmdefinition} can be directly compared. 

After determining the correlators $b_{lm}$,  we will incorporate the effects of the angular and energy warps to the spectrum. Following \cite{Abramo:2010gk}, we express the temperature $\Trec (\vec{P})$ at any point $\vec{P} = \l R, \, \beta, \, \phi\r$ on $\Sigma$ by the expansion 
\begin{eqnarray}
\Trec(\vec{P}) & = & \int_{\Sigma} \frac{d^{3}k}{\l2 \pi\r^3} \, e^{\, i \vec{k}.\vec{P}} \, \tilde{T}_{\text{rec}}(\vec{k})
\end{eqnarray}
The Fourier components $\tilde{T}_{\text{rec}}(\vec{k})$ represent the power spectrum at recombination. Since the anisotropic curvature is small in the era preceding recombination,  the  $\tilde{T}_{\text{rec}}(\vec{k})$ are drawn from a statistically isotropic, gaussian distribution.  Writing the term $e^{\, i \vec{k}.\vec{P}}$ using spherical harmonics, we have
\begin{eqnarray}
\Trec(\vec{P}) & = & \int_{\Sigma} \frac{d^{3}k}{\l2 \pi\r^3} \,  \tilde{T}_{\text{rec}}(\vec{k}) \, \times \, 4 \pi \sum_{lm} i^l \, j_l \l k \,R\l \beta \r \r \,  Y^{*}_{lm} (\hat{k} ) \, Y_{lm} \l \beta, \phi \r
\label{Eqn:blmintegral}
\end{eqnarray}
where $j_l$ are the spherical bessel functions and $Y_{lm}$ are the spherical harmonics. Using the expression for $R$ in equation \eqref{Eqn:Rdefinition}, we expand  $R \l \beta \r$ for small $\Omega_{k_0}$. Comparing this expansion with the definition of the $b_{lm}$ in equation \eqref{Eqn:blmdefinition}, we have 
\begin{eqnarray}
\nonumber
b_{lm} & =  & \int_{\Sigma} \frac{d^{3}k}{\l2 \pi\r^3} \,  \tilde{T}_{\text{rec}}(\vec{k}) \, \times \, 4 \pi \, i^l \\ && \, \l j_l \, Y_{lm}^{*} + \Omega_{k_0} \l - \, d_{l-2} \, f^{l-2,m}_{+2} \, Y^{*}_{l - 2, m}\,  +  \, d_{l} \, f^{l,m}_{0} \, Y^{*}_{l, m}\,    - \, d_{l+2} \, f^{l+2,m}_{-2} \, Y^{*}_{l + 2, m}\r\r
\label{Eqn:blmexpression}
\end{eqnarray}
The details of this expansion, including the definitions of the coefficients $d_l$ and $f^{lm}$ can be found in Appendices \ref{AppendixBessel} and \ref{AppendixSpherical}. The $Y_{lm}$ in the above expression are all functions of the unit vector $\hat{k}$ in the integrand.  Armed with the expression \eqref{Eqn:blmexpression}, we compute the correlators to first order in $\Omega_{k_0}$.  Each $b_{lm}$ receives contributions from the spherical harmonics $Y_{lm}$ and $Y_{l\pm2, m}$. Consequently,  we expect non zero power in the auto correlation of each mode and correlation between modes separated by 2 units of angular momentum. These correlators are
\begin{eqnarray}
\langle b_{lm} \, b^{*}_{lm} \rangle & = & C_{l} \l 1 \, + \, \frac{16}{45} \, \sqrt{\frac{\pi}{5}} \, \Omega_{k_0}\, \Delta_l  \, f^{lm}_{0} \r \\
\langle b_{lm} \, b^{*}_{l+2, m} \rangle & = & \frac{8}{45} \, \sqrt{\frac{\pi}{5}} \, \Omega_{k_0} \, \l f^{l+2, m}_{-2} \, \Delta_{l+2} \, C_{l+2} \, + \, f^{lm}_{+2} \, \Delta_l \, C_l \r 
\label{Eqn:blmanswers}
\end{eqnarray}
where the coefficients $\Delta_l$  are $\OO\l 1 \r$ numbers with  a weak dependence on $l$. All other correlators vanish. We relegate the details of this calculation to Appendix \ref{AppendixBessel}. 

Let us now relate the coefficients $a_{lm}$ and $b_{lm}$.  The present day temperature  $T_0$ is given by \eqref{Eqn:CMBTemperature}. The relationship between $\beta$ and $\theta_0$ can be obtained from their respective definitions \eqref{Eqn:betadefinition} and \eqref{Eqn:thetadefinition}. This relationship is given by 
\begin{eqnarray}
\beta &  = & \theta_0 - \frac{\Omega_{k_0}}{15} \sin 2 \theta_0
\label{Eqn:betasolution}
\end{eqnarray}
Owing to the $O(2)$ symmetry in the $\l r, \phi \r$ plane, the angle $\phi$ is the same as the azimuthal angle $\phi_0$ used by O.  We use the above relation to expand $T_0$ to leading order in $\Omega_{k_0}$, obtaining
\begin{equation}
\label{eq-effects}
 T_{0}\l \theta_0,\phi_0 \r  =  \left(\frac{a_P^2 \, b_P}{a_0^2 \, b_0}\right)^{1/3} \l 
 T_\mathrm{rec}(\theta_0,\phi_0) -
\frac{8}{15} \, \sqrt{\frac{\pi}{5}} \, \Omega_{k_0} \, Y_{20}\, 
 T_\mathrm{rec}(\theta_0,\phi_0) \, - \, 
\frac{\Omega_{k_0}}{15} \,  \sin(2\theta_0) \, \partial_{\theta_0}  \, T_\mathrm{rec}(\theta_0,\phi_0) \r
\end{equation}
The second term in the above expression arises as a result of the differential red shift caused by the non-isotropic Hubble expansion \eqref{Eqn:CMBTemperature}, while the third time arises due to the warp between the angles $\theta_0$ and $\beta$ (as in equation \eqref{Eqn:betasolution}).  This expansion is valid for angular scales $l \lessapprox \l \Omega_k \r^{-1}$. Using the spherical harmonic expansions for $T_0$ and $\Trec$ in terms of $a_{lm}$ and $b_{lm}$ respectively, we find 

\begin{equation} 
\label{eq-almblm}
a_{lm}= b_{lm}  - \Omega_{k_0} \, \l  h^{lm}_0 \,  b_{lm} + h^{l-2,m}_{+2}\,b_{l-2,m} + h^{l+2,m}_{-2}\, b_{l+2,m} \r
\end{equation}
The coefficients $h^{lm}$ in \eqref{eq-almblm} are obtained by combining the different spherical harmonics  in \eqref{eq-effects}. These coefficients are computed in the Appendix \ref{AppendixSpherical}. Using the correlators of the $b_{lm}$ (see equation \eqref{Eqn:blmanswers}), we can compute the expectation values 
\begin{eqnarray}
\nonumber
\langle a_{lm} \, a_{lm}^*\rangle & = & C_{l} \l 1 +  2 \, \Omega_{k_0} \l \frac{8}{45} \, \sqrt{\frac{\pi}{5}} \,  \Delta_l \, f^{lm}_0 \, + \, h^{lm}_0 \r \r \\ 
\langle a_{lm} \, a_{l+2,m}^*\rangle & = &   \Omega_{k_0} \, \l C_{l+2} \, \l \frac{8}{45} \, \sqrt{\frac{\pi}{5}} \,  \Delta_{l+2} \, f^{l+2,m}_{-2} \, + \, h^{l+2,m}_{-2} \r \, + \, C_{l} \,  \l \frac{8}{45} \, \sqrt{\frac{\pi}{5}} \,  \Delta_l \, f^{lm}_{+2} \, + \, h^{lm}_{+2} \r\r
\label{Eqn:almcorrelations}
\end{eqnarray}

Other correlation functions are unaffected by the anisotropic curvature $\Omega_{k_0}$.  Equation \eqref{Eqn:almcorrelations} specifies that modes separated by 2 units of angular momentum $l$ are mixed while there is no mixing between modes of different $m$. Physically, this implies correlations between modes of different angular scales (separated by two units of scale), but not of different orientation. The absence of mixing between modes of different orientation is due to the fact that the space-time preserves an $O(2)$ symmetry in the $\l r, \phi \r$ plane. However, even though there is no correlation between modes of different $m$, the power $\langle a_{lm} \, a_{lm}^{*} \rangle$ in a mode depends upon $m$ through the coefficients $f^{lm}_0$ and $h^{lm}_0$. Both these coefficients scale as $\sim \l l^2 - m^2\r$ (see equations \eqref{Eqn:flmdefinitionAppendix} and \eqref{Eqn:hlmdefinitionAppendix}). Hence, we expect different amounts of power in the high $m$ mode versus th
 e low $m$ mode for a given $l$. 

Equipped with the knowledge of the correlators \eqref{Eqn:almcorrelations}, we can perform tests of the statistical isotropy of the CMB. We follow the bipolar power spectrum analysis proposed by \cite{Hajian:2003qq} and adopt their notation (note that the normalization convention adopted by  \cite{Hajian:2003qq} is different from that used by the WMAP team \cite{Bennett:2010jb}). In this analysis, one computes the correlator
\begin{eqnarray}
A^{LM}_{ll^{'}} & = & \sum_{m m^{'}} \, \langle a_{lm} \,  a^{*}_{l'm'} \rangle \, \l -1\r^{m'} \mathcal{C}^{LM}_{l,m,l^{'},-m'}
\label{Eqn:SITest}
\end{eqnarray}
where the $\mathcal{C}^{LM}_{l,m,l^{'},-m^{'}}$ are Clebsch Gordan coefficients. In a statistically isotropic universe, these correlators are all zero except when $L = 0, M = 0$ and $l = l^{'}$. In the present case, we use the correlators \eqref{Eqn:almcorrelations} to compute the above statistic. For large $l$, the only non-zero correlators are
\begin{eqnarray}
\nonumber
A^{20}_{ll} & \approx & \l -1\r^{l} \,  \sqrt{l} \, \Omega_{k_0} \, \frac{2}{15} \,  \sqrt{\frac{2}{5}} \,  C_l  \, \l 1 \,  -  \, \frac{2}{3} \, \Delta_l \,  \r \\
A^{20}_{l + 2, l} & \approx &  \l -1\r^{l} \, \sqrt{l} \, \Omega_{k_0} \,  \frac{2}{15 \sqrt{15}} \, \l  l   \l C_{l+2} - C_{l}  \r + \l C_{l+2} \Delta_{l+2} + C_l \Delta_l \r  - \frac{15}{4} \l C_{l} - \frac{1}{5} C_{l+2} \r \r
\label{Eqn:SIAnswer}
\end{eqnarray}
Since the $C_l$ are smooth functions of $l$, $\l C_{l+2} - C_{l}  \r \sim \frac{C_l}{l}$. The above correlators then scale as 
\begin{equation}
A^{20}_{ll}  \sim A^{20}_{l, l+2}  \sim  \sqrt{l} \, \Omega_{k_0} \,  C_l 
\end{equation}
We note that these correlators are non-zero for all angular scales. This is precisely because the late time warp caused by the anisotropic curvature affects all the modes in the CMB. Consequently, this is a statistically robust test of anisotropy. Furthermore, this test of anisotropic curvature is immune to degeneracies from the expansion history of the universe that plague the measurement of isotropic curvature. Indeed, in an isotropic universe, irrespective of the cosmological expansion history, this statistic would be zero. 
This is similar to the effect discussed in section \ref{SubSec:Rulers} on standard rulers. In both cases, the anisotropic curvature affects measurements along every line of sight, breaking degeneracies with the cosmological expansion history. The similarity between these two effects is not surprising since the statistic \eqref{Eqn:SITest} captures the effect of the angular warp of the CMB by the anisotropic curvature (the third term in \eqref{eq-effects}). 

Minimum variance estimators obtained from the CMB temperature/polarization for a power asymmetry of this type and the observability are calculated in \cite{Pullen:2007tu}.  Statistical analyses of the sort discussed in this section have been performed with the WMAP data \cite{Bennett:2010jb}. In a universe with anisotropic curvature, these statistical tests can lead to quadrupolar dependence of the two point function. The expected answer for the statistic  \eqref{Eqn:SIAnswer} has power only in the $A^{20}_{ll}$ and $A^{20}_{l, l-2}$ modes. Furthermore, since these correlators are proportional to $C_l$, the effect shows a bump around the first acoustic peak. Interestingly, the two point quadrupolar anomaly in the WMAP data shows similar characteristics with power only in the $A^{20}_{ll}$ and $A^{20}_{l, l-2}$ modes, which peaks around the first acoustic peak.  This anomaly could be explained in our scenario if the anisotropic curvature $\Omega_ {k_0} \sim 10^{-2}$. However, such a large anisotropic curvature is heavily constrained by the absence of a correspondingly large quadrupole in the CMB (see  section \ref{SubSec:CMBQuadrupole}). While this  anomaly may be due to other systematic effects \cite{Bennett:2010jb}, similar searches could be performed with upcoming CMB experiments. It is conceivable that these experiments could discover correlations from anisotropic curvatures $\Omega_{k_0} \sim 10^{-4}$, as allowed by the size of the CMB quadrupole.

\subsection{Compact Topology}

We have so far considered only signals arising from the geometry of the universe, but observable signals may also arise from the topology.  The normal eternal inflation picture makes it appear that space should be very large or infinite in all directions \cite{Tegmark:2004qd}.  If our observable universe nucleated as a bubble from (3+1 dimensional) false vacuum inflation then it will appear as an infinite, open universe.  However, in our picture it is natural that the observable universe could have one or two compact dimensions, even though it came from an eternally inflating space \footnote{This assumes of course that the transition to our vacuum was not topology-changing.}.  Interestingly, the size of these compact dimensions may be close to the Hubble scale today because the period of slow-roll inflation was not too long.  In the case of a 2+1 dimensional parent vacuum, the topology of the spatial dimensions of the observable universe would be $\mathbb{R}^2 \times S^1$.  Since the curvature is all in the $\mathbb{R}^2$ and not the $S^1$, the curvature radius of the universe and the topology scale (in this case the radius of the $S^1$) are disconnected.  Thus, even though the curvature radius today is restricted to be $\sim 10^2$ times longer than the Hubble scale, the size of the compact dimension can be smaller than the Hubble scale.  In fact, we expect that slow-roll inflation began when the curvature radius was around the Hubble scale of inflation.  Thus, for the $S^1$ to be around the Hubble scale today it would have needed to be about $10^2$ times smaller than the Hubble size at the beginning of inflation.  For high scale inflation this is near the GUT scale, a very believable initial size for that dimension.  This scenario is interestingly different from the compact topologies often considered, for which an isotropic geometry ($S^3$, $E^3$ or $H^3$) is usually assumed (though see \cite{Mota:2003jb}).  Any compact topology necessarily introduces a global anisotropy, but in our scenario even the local geometry of the universe is anisotropic.  This allows the curvature radius and the topology scale to be different by orders of magnitude.

Thus it is reasonable that in our picture we may also have the ``circles in the sky" signal of compact topology \cite{Cornish:1996kv}.  Current limits from the WMAP data require the topology scale to be greater than 24 Gpc \cite{Cornish:2003db}.  This limit can be improved by further searching, especially with data from the Planck satellite, to close to the $\sim 28$ Gpc diameter of our observable universe.  If discovered in conjunction with anisotropic curvature this would provide a dramatic further piece of evidence that we originated in a lower dimensional vacuum.  Further the directions should be correlated.  If the parent vacuum was 2+1 dimensional then we expect the circles in the sky to be in the previously compact direction (the $S^1$) while the curvature is in the other two dimensions.  On the other hand, if the parent vacuum was 1+1 dimensional then it seems possible that both the signals of curvature and the compact topology would be in the same two dimensions, with the third dimension appearing flat and infinite.  Thus seeing both the anisotropic curvature and signals of the compact topology may provide another handle for determining the dimensionality of our parent vacuum.

\subsection{Other Measurements}
The CMB is a precise tool to measure cosmological parameters. However, it is a two dimensional snapshot of the universe at a given instant in time.  Additional information can be obtained through three dimensional probes of the universe.  Several experiments that yield three dimensional data  are currently being planned. These include 21 cm tomography experiments and galaxy surveys. A complete study of the effects of anisotropic curvature in these experiments is beyond the scope of this work. In this section, we briefly mention some  possible tests of this scenario in these upcoming experiments.  

A three dimensional map of the universe can be used to distinguish anisotropic curvature  from fluctuations in the matter density. Anisotropic curvature does not lead to inhomogeneities in the matter distribution. Consequently, measurements of the large scale matter density can be used to distinguish between these two situations. Such measurements may be possible using upcoming 21 cm experiments and high redshift surveys, for example LSST.  LSST should be sensitive to isotropic curvatures down to $\sim 10^{-3}$ with objects identified out to redshift $z \approx 1$ \cite{Ivezic:2008fe}.  Since the dominant effect of anisotropic curvature occurs at late times, LSST should be a good way to probe our signals. Additionally, 21 cm experiments may also be sensitive to isotropic curvatures $\Omega_{k_0} \sim 10^{-4}$ \cite{Mao:2008ug}, and so may offer a very precise test of anisotropic curvature.

The curvature anisotropy also gives rise to a differential Hubble expansion rate  $\Delta H \sim \Omega_{k_0} \, H_a$ (see Section \ref{Sec: metric}),  which contributes to the quadrupole in the CMB  (see section \ref{SubSec:CMB}). This effect will also be visible in direct measurements of the Hubble parameter. Current experimental constraints on this effect are at the level of a few percent \cite{Rubin} and are likely to become better than $\lessapprox 10^{-2}$ in future experiments  \cite{Schutz:2001re, MacLeod:2007jd}.

\section{Discussion}
\label{Sec: Conclusions}

A universe produced as a result of bubble nucleation from an ancestor vacuum which has two large dimensions and one small, compact dimension is endowed with anisotropic curvature $\Omega_k$. Such an anisotropic universe is also produced in the case when our 3+1 dimensional universe emerges from a transition from a 1+1 dimensional vacuum. In this case, depending upon the curvature of the compact dimensions, the resulting universe can have either positive or negative curvature along two dimensions, with the other remaining flat. The geometry of the equal time slices of the daughter universe are such that two of the directions are curved while the other dimension is flat. Immediately after the tunneling event, the energy density of the universe is dominated by this anisotropic curvature $\Omega_k$. This curvature drives the  curved directions to expand differently from the flat direction, resulting in  differential Hubble expansion $\Delta H$ between them.

The expansion of the universe dilutes $\Omega_k$ until it becomes small enough to allow slow roll inflation. At this time, the universe undergoes a period of inflation during which the curvature $\Omega_k$ and the differential Hubble expansion $\Delta H$ are exponentially diluted. However, during the epochs of radiation and matter domination, the curvature red shifts less strongly than either the radiation or the matter density. Consequently, the fractional energy density $\Omega_k$ in curvature grows with time during these epochs. This late time emergence of an anisotropic curvature $\Omega_k$ also drives a late time differential Hubble expansion $\Delta H$ in the universe.

These late time, anisotropic warps of the space-time geometry are all proportional to the current fractional energy density in curvature,  $\Omega_{k_0}$. They can be observed in the present epoch if inflation does not last much longer than the minimum number of efolds required to achieve a sufficiently flat universe ($\sim 65$ efolds for high scale inflation). Anisotropic curvature leads to the warping of the angular size of standard rulers. This warping is a function of both the angle and orientation of the ruler in the sky. Consequently, this effect is immune to degeneracies from the expansion history of the universe since it affects rulers that are along the same line of sight but oriented differently. 

The CMB is also warped by the anisotropic curvature. In addition to the geometric warping, the differential Hubble expansion $\Delta H$ also preferentially red shifts the energies of the CMB photons. This energy shift differentially changes the monopole temperature of the CMB giving rise to a quadrupole in the CMB. Furthermore, since the anisotropic curvature is a late time effect, it affects all the modes that can be seen in the CMB. Consequently, this effect leads to statistical anisotropy on all angular scales. This effect is different from other signatures of anisotropy considered in the literature  \cite{Gumrukcuoglu:2007bx, Gumrukcuoglu:2008gi}. Previous work has concentrated on the correlations that are produced due to the initial anisotropy in the universe at the beginning of inflation. Since these modes are roughly stretched to the Hubble size today, these initial anisotropies only affect the largest modes in the sky and are hence low l effects in the CMB. The late time anisotropy however warps the entire sky and leads to statistically robust correlations on all angular scales. The anisotropies in the pre-inflationary vacuum can however lead to other interesting signatures, for example in the gravitational wave spectrum \cite{Gumrukcuoglu:2008gi}. These signatures are an independent check of this scenario. Anisotropies that affect all angular scales have also been previously considered \cite{Ackerman:2007nb, Boehmer:2007ut}. These required violations of rotational invariance during inflation and the anisotropy emerges directly in the primordial density perturbations. In our case, the density perturbations are isotropic and the anisotropy observed today is a result of a late time  warp of the space-time. 

Anisotropic curvature is already more stringently constrained than isotropic curvature. While isotropic curvature is bounded to be $\lessapprox 10^{-2}$, it is difficult for anisotropic curvature to be much larger than $\sim 10^{-4}$ without running afoul of current data, in particular, the size of the CMB quadrupole. Since the measurement of curvature is  ultimately limited by cosmic variance $\sim 10^{-5}$, there is a window between $ 10^{-5} \lessapprox \Omega_{k_0} \lessapprox 10^{-4}$ that can be probed by upcoming experiments, including Planck. 

Future cosmological measurements like the 21 cm experiments will significantly improve bounds on the curvature of the universe.  A discovery of isotropic curvature would be evidence suggesting that our ancestor vacuum had at least three large space dimensions. On the other hand, a discovery of anisotropic curvature would be strong evidence for the lower dimensionality of our parent vacuum. The anisotropy produced from such a transition has a very specific form due to the symmetries of the transition. It leads to correlations only amongst certain modes in the CMB (for example, only $A^{20}_{ll}$ and $A^{20}_{l, l-2}$). This distinguishes it from a generic anisotropic $3+1$ dimensional pre-inflationary vacuum which will generically have power in all modes. In these scenarios, it is also natural for the universe to have non-trivial topology. The existence of a non-trivial topological scale within our observable universe can be searched for using the classic ``circles 
 in the sky" signal.  If both the non-trivial topology and anisotropic curvature can be discovered, implying a period of  inflation very close to the catastrophic boundary, it would be powerful evidence for a lower dimensional ancestor vacuum.  A discovery of these effects would establish the existence of vacua vastly different from our own Standard Model vacuum, lending observational credence to the landscape.

%technical summary of our universe expansion history + observables

%para about nontrivial topology

\begin{comment}
- the delta H comes back at late times just like curvature
- we have a high-l observable, not just low-l
	these guys \cite{Gumrukcuoglu:2007bx} compute the correlators Alm in an initially anisotropic universe but they only find an effect for the lowest l's precisely because they don't have our anisotropic curvature so don't have our late time effect!
- this anisotropic signal is a different way of probing curvature, not subject to the usual degeneracy with expansion history

- there may also be interesting gravity wave signatures of our scenario \cite{Gumrukcuoglu:2008gi}

- even if curvature is observed at $10^-2$ (and our signal not seen) then we know ancestor vacuum had 3 or more large spatial dimensions
\end{comment}

\section*{Acknowledgments}
We would like to thank Savas Dimopoulos, Sergei Dubovsky, Ben Freivogel, Steve Kahn, John March-Russell, Stephen Shenker, Leonard Susskind, and Kirsten Wickelgren for useful discussions and the Dalitz Institute at Oxford for hospitality.  S.R. was supported by the DOE Office of Nuclear Physics under grant DE-FG02-94ER40818.  S.R. is also supported by NSF grant PHY-0600465.

While this work was in progress we became aware of interesting work by another group working on different signals of a similar general framework \cite{BlancoPillado:2010uw}.

\appendix
\section{Calculation of the Correlations}
\label{AppendixBessel}
The temperature  $\Trec (\vec{P})$  at any point $\vec{P} = \l R, \, \beta, \, \phi\r$ on the surface of last scattering $\Sigma$ (see figure \ref{Fig:geodesics}) can be expressed using spherical harmonics (see equation \eqref{Eqn:blmintegral})
\begin{eqnarray}
\Trec(\vec{P}) & = & \int_{\Sigma} \frac{d^{3}k}{\l2 \pi\r^3} \,  \tilde{T}_{\text{rec}}(\vec{k}) \, \times \, 4 \pi \sum_{lm} i^l \, j_l \l k \,R\l \beta \r \r \,  Y^{*}_{lm} (\hat{k} ) \, Y_{lm} \l \beta, \phi \r
\label{Eqn:blmintegralAppendix}
\end{eqnarray}
Expanding the Bessel junctions $j_l$ in \eqref{Eqn:blmintegralAppendix} around $R_0 \, = \, 3 \, t_0^{\frac{1}{3}} \, t_r^{\frac{2}{3}} \, \l 1 \, + \, \frac{\Omega_{k_0}}{45} \r $ to linear order in $\Omega_{k_0}$, we get 
\begin{eqnarray}
j_l \l k R \l \beta \r \r & = & j_l \l k R_0 \r \, + \, \Omega_{k_0} \, d_l \, Y_{20} \l \beta, \phi \r
\label{Eqn:BesselExpansion}
\end{eqnarray}
where the coefficient $d_l$ is 
\begin{eqnarray}
d_l & = &  \frac{8}{45} \, \sqrt{\frac{\pi}{5}} \, \l  k R_0 \, j_{l+1} \l k R_0\r - l \, j_l \l k R_0\r\r
\label{Eqn:dldefinition}
 \end{eqnarray}
This expansion is valid for $l \lessapprox \l \Omega_{k_0} \r^{-1}$. The spherical harmonic $Y_{20}\l \beta, \phi \r$ in  \eqref{Eqn:BesselExpansion} multiplies $Y_{lm} \l \beta, \phi \r $ in the expansion \eqref{Eqn:blmintegralAppendix}. These harmonics can be combined, yielding
\begin{eqnarray}
Y_{20} \, Y_{lm} & = & f^{lm}_{-2} \,  Y_{l-2, m} \, + \,  f^{lm}_{0} \,  Y_{l, m} \, +  \, f^{lm}_{+2} \,  Y_{l+2, m} 
\label{Eqn:Y20YlmAppendixA}
\end{eqnarray}
The definitions of the $f^{lm}$ are given in Appendix \ref{AppendixSpherical}. Using  \eqref{Eqn:Y20YlmAppendixA}, the coefficient $b_{lm}$ of $Y_{lm} \l  \beta, \phi \r$  in \eqref{Eqn:BesselExpansion} is  the expression  in equation \eqref{Eqn:blmexpression}. With this information, we can compute the correlations amongst the $b_{lm}$. Imposing the requirement that $\tilde{T}_{\text{rec}}(\vec{k})$ are drawn from a statistically isotropic, gaussian distribution \cite{Abramo:2010gk},  the two point function $\langle b_{lm} \, b^{*}_{lm} \rangle$  to linear order in $\Omega_{k_0}$ is
\begin{eqnarray}
\langle b_{lm} \, b^{*}_{lm} \rangle & = & \int \frac{dk}{k} \, \frac{2}{\pi} \, N_2 \l k \r \l j_l^2 \, + \, 2 \, \Omega_{k_0} \, j_l \, d_l  \, f^{lm}_0\r
\label{Eqn:blmcorrelatorsAppendix}
\end{eqnarray}
where $N_2 \l k \r$ is the  two point function $ \left(\frac{a_P^2 \, b_P}{a_0^2 \, b_0}\right)^{2/3} \langle \Delta\Trec (\vec{k}) \, \Delta\Trec(\vec{k}) \rangle$  of the temperature anisotropies $ \Delta\Trec $ (as defined in \cite{Abramo:2010gk}). We have again scaled out the piece that accounts for the red shift between the era of recombination and the present epoch. The first term in the integrand is the usual contribution $C_l$  to the power in the $l$ mode. The second term, proportional to $\Omega_{k_0}$, arises from the anisotropic curvature. To compute this term, we substitute for $d_l$ (from \eqref{Eqn:dldefinition}) in \eqref{Eqn:blmcorrelatorsAppendix}. The resulting integral has the form 
\begin{equation}
 \int \frac{dk}{k} \, \frac{2}{\pi} \, N_2\l k \r \,  j_l \,  \l k R_0 \, j_{l+1} \, -  \, l \, j_l  \r
\label{Eqn:dlintegral}
\end{equation}
The second term in the above integrand is  $l \, C_l$. For the first term, 
\begin{eqnarray}
S_l & = & \int \frac{dk}{k} \, \frac{2}{\pi} \, N_2\l k \r \,  kR_0 \, j_l \,  j_{l+1} 
\label{Eqn:Sldefinition}
\end{eqnarray}
we use the fact that the $j_l$ satisfy the identity 
\begin{equation}
j_{l+1} \l kR_0\r \, + \, j_{l-1}\l kR_0\r \, = \, \frac{2 l \, + \, 1}{kR_0} \, j_l \l kR_0\r 
\end{equation}
This implies 
\begin{eqnarray}
S_l \, + \, S_{l-1} & = & \l 2 l \, + 1 \r C_l 
\end{eqnarray}
Physically, since there is roughly similar amounts of power in all the $C_l$, we expect $S_l \sim S_{l-1}$. This implies 
\begin{eqnarray}
S_l & = & l \,  C_l  \, + \, \Delta_l \, C_l 
\label{Eqn:Slsolution}
\end{eqnarray}
where $\Delta_l $ is an order one coefficient. $\Delta_l$ can be computed by integrating \eqref{Eqn:Sldefinition}.  This calculation requires explicit use of the two point function $N_2 \l k \r$ at recombination and is beyond the scope of this paper. 

Using \eqref{Eqn:Slsolution} in \eqref{Eqn:dlintegral} and \eqref{Eqn:blmcorrelatorsAppendix} , we have 
\begin{eqnarray}
\langle b_{lm} \, b^{*}_{lm} \rangle & = & C_l \, \l 1 + \frac{16}{45} \, \sqrt{\frac{\pi}{5}} \, \Omega_{k_0} \, \Delta_l \, f^{lm}_0\r 
\end{eqnarray}

A similar calculation can be performed for the other correlators of the $b_{lm}$. In the expression \eqref{Eqn:blmexpression} for $b_{lm}$,  each $b_{lm}$ receives contributions from the spherical harmonics $Y_{lm}$ and $Y_{l\pm2, m}$. Consequently, we expect non trivial correlations only between modes separated by 2 units of angular momentum. This correlator is 
\begin{eqnarray}
\langle b_{lm} \, b^{*}_{l+2, m} \rangle & = & \frac{8}{45} \, \sqrt{\frac{\pi}{5}} \, \Omega_{k_0} \, \l f^{l+2, m}_{-2} \, \Delta_{l+2} \, C_{l+2} \, + \, f^{lm}_{+2} \, \Delta_l \, C_l \r 
\end{eqnarray}

With this knowledge, we can compute the correlators of the $a_{lm}$. Equation \eqref{eq-effects} expresses the temperature $T_0$ (characterized by $a_{lm}$) observed today in terms of the temperature $\Trec$ (characterized by $b_{lm}$)  at recombination. Writing $\Trec$ in terms of the $b_{lm}$ in  \eqref{eq-effects}, we get, 
 \begin{eqnarray}
 T_0 \l \theta_0, \phi_0 \r & = & \sum_{lm} \l b_{lm} \l  Y_{lm} \, - \, \Omega_{k_0} \, \l  \frac{8}{15} \, \sqrt{\frac{\pi}{5}} \,  Y_{20}  \, Y_{lm} \,  + \, \frac{1}{15}  \,\sin \l 2 \theta_0 \r \,  \partial_{\theta_0} \, Y_{lm} \r \r \r
 \label{Eqn:T0expressionAppendix}
\end{eqnarray}
The spherical harmonics in \eqref{Eqn:T0expressionAppendix} are all functions of $\l \theta_0, \phi_0 \r$. The products of the spherical harmonics $Y_{20} \, Y_{lm}$ and $\sin \l 2 \theta_0 \r \partial_{\theta_0} \, Y_{lm}$ can be expressed as a linear combination of the $Y_{lm}$. The term $Y_{20} \, Y_{lm}$ can be expressed as the combination \eqref{Eqn:Y20YlmAppendixA}, whilst  $\sin \l 2 \theta_0 \r \partial_{\theta_0} \, Y_{lm}$ is expressed as 
\begin{eqnarray}
\sin \l 2 \theta_0 \r \partial_{\theta_0} \, Y_{lm} & = & g^{lm}_{-2} \, Y_{l-2, m} \, + \, g^{lm}_{0} \, Y_{lm} \, + \, g^{lm}_{+2} \, Y_{l+2, m}
\label{Eqn:Y20PartialYlm}
\end{eqnarray}
The coefficients $g^{lm}$ are defined in Appendix \ref{AppendixSpherical}.  Using  \eqref{Eqn:Y20YlmAppendixA} and \eqref{Eqn:Y20PartialYlm} in \eqref{Eqn:T0expressionAppendix}, we get the expression \eqref{eq-almblm} for the $a_{lm}$ in terms of the $b_{lm}$. The $a_{lm}$ in \eqref{eq-almblm} are expressed as a linear combination of $b_{lm}$ and $b_{l \pm 2, m}$. Consequently, to linear order in $\Omega_{k_0}$, we expect power in the modes $\langle a_{lm} \, a^{*}_{lm} \rangle$ and  $\langle a_{lm} \, a^*_{l\pm2, m} \rangle$. Using  \eqref{eq-almblm}, the correlators \eqref{Eqn:almcorrelations} can be computed. 

We now give the exact answers for the measures of statistical anisotropy computed approximately in Eqn.~\eqref{Eqn:SIAnswer}:
\begin{eqnarray}
A^{20}_{l, l} & = & -\frac{4 (-1)^l l \left(1+3 l+2 l^2\right) \Omega_{k_0} C_l (-3+2 \Delta_l)}{45 \sqrt{5} \sqrt{l \left(-3-5 l+10 l^2+20 l^3+8 l^4\right)}} \\
A^{20}_{l + 2, l} & = & -\frac{2 (-1)^l \sqrt{\frac{2}{15}} (1+l) (2+l) \Omega_{k_0} (C_l (3+l-\Delta_l)-C_{l+2} (l+\Delta_{l+2}))}{15 \sqrt{6+13 l+9 l^2+2 l^3}}
\end{eqnarray}

\section{Spherical Harmonics}
\label{AppendixSpherical}
In this Appendix, we give the definitions of the coefficients $f^{lm}$, $g^{lm}$ and $h^{lm}$. 

The $f^{lm}$ are defined by the relation 
\begin{eqnarray}
\nonumber
Y_{20} \, Y_{lm} & = & f^{lm}_{-2} \,  Y_{l-2, m} \,  + \, f^{lm}_{0} \,  Y_{lm}  \, + \, f^{l+2, m}_{+2} \, Y_{l+2, m}
\end{eqnarray}
They evaluate to 
\begin{eqnarray}
\nonumber
f^{lm}_{-2} & = &\frac{3 \sqrt{(2 l-3) (2 l+1) (l-m-1) (l-m) (l+m-1) (l+m)} \sqrt{\frac{5}{\pi }}}{4 \left(8 l^3-12 l^2-2 l+3\right)} \\
\nonumber
f^{lm}_0 & = & \frac{\left(l^2+l-3 m^2\right) \sqrt{\frac{5}{\pi }}}{8 l (l+1)-6} \\
f^{lm}_{+2} & = & \frac{3 \sqrt{(2 l+1) (2 l+5) (l-m+1) (l-m+2) (l+m+1) (l+m+2)} \sqrt{\frac{5}{\pi }}}{4\left(8 l^3+36 l^2+46 l+15\right)}
\label{Eqn:flmdefinitionAppendix}
\end{eqnarray}

The $g^{lm}$ are defined by the relation 
\begin{eqnarray}
\nonumber
\sin \l 2 \theta_0 \r \partial_{\theta_0} \, Y_{lm} & = & g^{lm}_{-2} \, Y_{l-2, m} \, + \, g^{lm}_{0} \, Y_{lm} \, + \, g^{lm}_{+2} \, Y_{l+2, m}
\end{eqnarray}
They evaluate to 
\begin{eqnarray}
\nonumber
g^{lm}_{-2} & = & -\frac{2 (l+1) \sqrt{\frac{(l-m) (l+m) \left((l-1)^2-m^2\right)}{4 l^2-4 l-3}}}{2 l-1} \\
\nonumber
g^{lm}_{0} & = & \frac{6 m^2-2 l (l+1)}{4 l (l+1)-3} \\
g^{lm}_{+2} & = & \frac{2 l \sqrt{(2 l+1) (2 l+5) (l-m+1) (l-m+2) (l+m+1) (l+m+2)}}{8 l^3+36 l^2+46 l+15}
\end{eqnarray}

The coefficients $h^{lm}$ (see equation \eqref{eq-almblm})are defined by the addition of the effects from the energy warp of the CMB by the anisotropic Hubble expansion and the warp of the angle $\theta_0$ at which the photon is observed and the angle $\beta$ at which it was emitted at the surface of last scattering. These evaluate to  
\begin{eqnarray}
\nonumber
h^{lm}_{-2} & = & \frac{2 (l-2) \sqrt{\frac{(l-m) (l+m) \left((l-1)^2-m^2\right)}{4 l^2-4 l-3}}}{30 l-15} \\
\nonumber
h^{lm}_{0} & = & -\frac{2 \left(l^2+l-3 m^2\right)}{15 (4 l (l+1)-3)}  \\
h^{lm}_{+2} & = & -\frac{2 (l+3) \sqrt{(2 l+1) (2 l+5) (l-m+1) (l-m+2) (l+m+1) (l+m+2)}  }{15 \left(8l^3+36 l^2+46 l+15\right)}
\label{Eqn:hlmdefinitionAppendix}
\end{eqnarray}

\end{document}